# A Highly Efficient Polarization-Independent Metamaterial-Based RF Energy-Harvesting Rectenna for Low-Power Applications

C. Fowler (cfowler5@mail.usf.edu)   Department of Physics, University of South Florida, 4202 E. Fowler Avenue, Tampa, FL 33620

J. Zhou (jiangfengz@usf.edu)   Department of Physics, University of South Florida, 4202 E. Fowler Avenue, Tampa, FL 33620

**Abstract**

A highly-efficient multi-resonant RF energy-harvesting rectenna based on a metamaterial perfect absorber featuring closely-spaced polarization-independent absorption modes is presented. Its effective area is larger than its physical area, and so efficiencies of 230% and 130% are measured at power densities of 10 µW/cm$^2$ and 1 µW/cm$^2$ respectively, for a linear absorption mode at 0.75 GHz. The rectenna exhibits a broad polarization-independent region between 1.4 GHz and 1.7 GHz with maximum efficiencies of 167% and 36% for those same power densities. Additionally, by adjustment of the distance between the rectenna and a reflecting ground plane, the absorption frequency can be adjusted to a limited extent within the polarization-independent region. Lastly, the rectenna should be capable of delivering 100 µW of power to a device located within 50 m of a cell-phone tower under ideal conditions.

**Introduction**

Although batteries have facilitated the wide-spread adoption of numerous portable electronic devices, they are not an ideal solution for implementation when costs, recharging, and/or large numbers of devices make their use prohibitive, as is often the case for various types of sensor networks[2]. An alternative approach with the potential to be more economical and convenient is by means of capturing ambient RF signals with an antenna and converting the induced electrical currents to DC power with a rectification system. To be practical, such an energy harvesting device should be highly efficient, compact in size, possess large bandwidth, and be polarization-independent. With antennas there is a tradeoff between the maximum power captured and directionality[4] that will also need to be taken into consideration with any design.

Historically, this approach has yet to be commercially successful due to the low amounts of ambient energy available[5-9], and the reduced efficiency of rectification systems when operating under low-power conditions[10]. Indeed, there is reason to believe that this approach will never be successful since there is simply not enough power available to be harvested for operating devices with even minimal power requirements (≈100 µW)[11]. Nonetheless, there are at least three reasons for the continued pursuit of ambient RF energy harvesting technology: 1. The increasing use of Wi-Fi networks, Cell-phones, Blue Tooth Devices, and so forth, may increase the amount of ambient power available to a level large enough for wireless RF energy harvesting to be practical. 2. The technology can potentially be scaled to regions of the electromagnetic spectrum where ambient power is more plentiful, such as infrared and optical. 3. The technology can easily be adopted for wireless power transport, where the power is collected from a dedicated source rather than from ambient sources.

While conventional antennas have never been adequate to capture enough RF power from ambient sources alone to be effective, the invention of metamaterial perfect absorbers (MPA)[12,13], has re-opened the possibility of practical ambient RF energy harvesting[14-25]. Metamaterials are composed of a 2-D or 3-D array of resonating structures. The arrangement and design of the structures allows the optical properties of the bulk material, such as the electric permittivity and magnetic permeability[26], to be tuned to a desired value. This effect primarily occurs at the resonance frequency of the structure. A metamaterial perfect absorber is a material where the imaginary components of the permittivity and permeability are maximized, while the real components are tuned to be impedance-matched to free space so that reflection is eliminated and electromagnetic waves at the resonance frequency are completely absorbed in the material. A perfect absorber is composed of a 2-D array of resonating structures along with an electrically conducting ground plane placed parallel to the array and separated by a substrate of appropriate thickness.

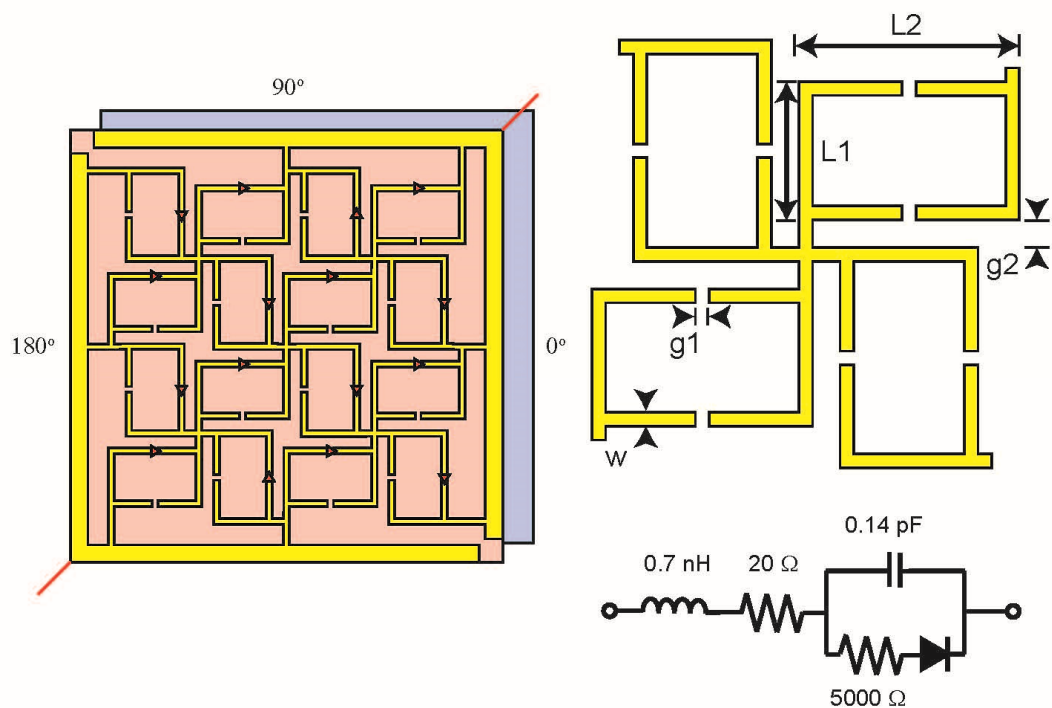

*Figure 1: Design of polarization-independent energy harvesting rectenna with polarization angles indicated along with the equivalent circuit model for Schottky diodes used for simulations. The direction and placement of Schottky diodes is indicated by red triangles. The parameters for the diodes were pulled from the specifications sheet. The red lines at the corners indicate where the connections to the load are made.  L1 =10 mm, L2 = 16 mm, g1 = 1 mm, w = 1 mm, g2 = 2 mm. The total area is about 54 cm².*

If a rectification system is built into a perfect absorber, the power absorbed by the material can be harvested and used or stored instead of lost. Such an RF energy harvesting device can potentially be constructed to be highly efficient, polarization-independent, electrically small[27,28], and omnidirectional. In this work, we present a design built from a network or split ring resonators that exhibits high efficiency at low power, good polarization-independence, limited tunability, and a relatively small physical area.

## **Method**

The design was made in *CST Microwave Studio* and simulated using a plane wave and the transient solver. The efficacy of the design was determined by using far field monitors to calculate the broadband absorption cross section (ACS) and by monitoring the power delivered to an impedance-matched load. The ACS, defined as the power absorbed by the sample divided by the incident power density, helps identify the presence and location of resonance peaks that result in optimal absorption, while the power measurement indicates how strongly the captured RF power is converted to DC power. The diodes were modelled using an equivalent circuit (figure 1) based from some found in the literature[29,30]. Various angles and ground plane distances were scanned to characterize the performance of the devices.

After adjusting parameters and optimizing the design, a physical sample was constructed by the use of conventional photolithography techniques and then afterwards soldering diodes onto the surface in the arrangement indicated in figure 1. The samples are tested using a typical system for transmitting a sinusoidal signal composed of a horn antenna, signal generator, amplifier, signal analyzer, and digital multimeter. Ideally, the measurements would take place in an anechoic chamber to reduce multipath interference. That has been forgone here for simplicity, but it introduced some fluctuations in measurements. A decade resistor box is used as a proxy for a device, and is connected across the terminals of the rectenna. The optimum value of the load resistance for maximum power transfer can be found by adjusting the resistor box. The efficiency ($\eta$) is defined as the power delivered to the load divided by the power density, S, and the geometric area of the sample (equations (1) and (2))[2]

$$\eta = \frac{P_{Load,DC}}{S} \times 100\% = \frac{V_{Load,DC}^2}{R_{Load} \times S} \times 100\% \tag{1}$$

$$S = \frac{GP_{out}}{4\pi d^2} \tag{2}$$

$P_{Load,DC}$ is the DC power delivered to the load, $G$ is the gain of the horn antenna, $P_{out}$ is the power delivered to the horn antenna, and $d$ is the distance between the horn antenna and the energy harvester. This definition can lead to efficiency measurements greater than 100% as also noted by Alavikia, Almoneef, and Ramahi[31], which does not mean that the rectenna delivers more power to the load than it receives. Rather, it means that the effective area of the rectenna is larger than its physical area, which is possible for non-aperture antennas (a dipole antenna is an example).

The first step in testing the samples was to find the optimum resistance of the load for delivering power. Due to the nonlinear nature of the diodes, the optimum load value is a function of the input power and the frequency. However, experiments with the sample indicate that the deviations only seem to affect the efficiency by around 3 or 4 percentage points provided that there is enough input power to activate the diodes, and so finding the optimum load at a single frequency with a reasonable power level ($\sim 10 \, \mu W/cm^2$) will work satisfactorily. Then the efficiency is measured while varying the angle of incidence in 15 degree steps over a 180 degree arc (only 180 degrees is necessary, because the oscillatory nature of the sinusoidal electromagnetic fields results in 180 degree rotational symmetry) and sweeping the frequency over the full range of our system (0.7-2.0 GHz) in 10 MHz steps, while keeping the transmitted power density constant at 10 μW/cm². This allowed the resonance peaks to be identified and the polarization dependence to be determined. A conducting metal plate with identical area to the sample is then placed behind the samples to create a Fabry-Perot cavity. Adjusting the distance between the plate and the sample allows for selectively enhancing specific peaks by matching them up with the resonance modes of the cavity. After placing the ground plane at the optimal distance for a specified resonance peak, the sample is again measured through another 180 degree arc to determine the enhancement. As an additional means of characterizing the design, the minimal power density required to deliver 100 μW to a load is indicated for each polarization angle.

## Results

After making a quick sweep to find resonance peaks, the optimum load was found to be 4000 Ω at an absorption peak found near 0.75 GHz with a 10 µW/cm$^2$ power density. Before the ground plane is introduced, the efficiency of the sample behaves as shown in figures 2a and 2b. Both the experimental and simulation results show a strong absorption peak near 0.75 GHz that, although highly efficient (≈230%), does not exhibit polarization-independence but instead exhibits dipole-like behavior. Additional resonances are more readily apparent in the simulation results, than in the experimental results. When a Fabry-Perot cavity is created by introducing the ground plane, these additional resonances manifest much more strongly, particularly when the cavity resonance frequency matches the absorption resonance frequency. When the cavity length is 2-4 cm, a broad polarization-independent region emerges with efficiency up to 167% as shown in figures 2c and 2d. The presence of the ground plane actually decreases the efficiency of the absorption peak at 0.75 GHz, which would otherwise be superior, but the efficiency is still fairly large (≈140%) when the ground plane is placed at the optimal distance for the polarization-independent region (30 mm).

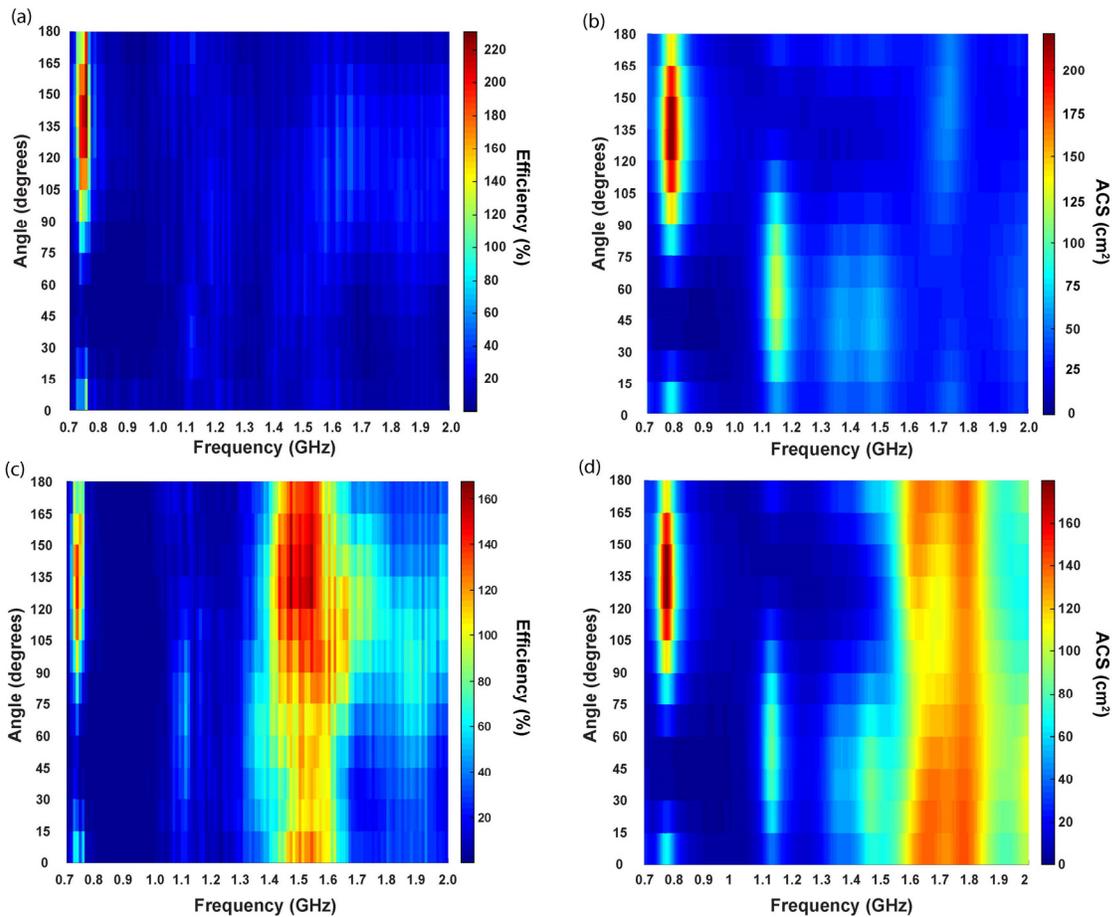

*Figure 2: Polarization dependence of energy harvesting efficiency. Experimental measurements (a) and ACS simulations (b) without the ground plane. Experimental measurements (c) and ACS simulations (d) with the ground plane spaced 30 mm behind the rectenna.*

Figure 3 shows the effect that the ground plane distance has on absorption and efficiency at 45 and 135 degree polarization angles. Of particular note is that the location of the polarization-independent region shifts as the ground plane is adjusted, demonstrating tunability in addition to its relatively broad range of absorption frequencies. Unfortunately, the quality of polarization-independence breaks down when the ground plane distance becomes larger than about 4 cm. The broadband behavior is caused by the presence of multiple overlapping absorption peaks in the region rather than by a single broad peak, which also explains the shift in peak absorption frequency as the ground plane is adjusted. Lastly, at 45 degrees, an additional dipole-like absorption peak emerges near 1.1 GHz. This peak reaches its maximum efficiency (≈99%) at a 45 degree polarization angle with a cavity length of 8 cm. It is the weakest absorption peak and is not polarization-independent.

Comparison of the experimental results with the simulations indicate some of the weaknesses of the simulations. While the ACS does a fairly good job of predicting where resonance peaks occur, it is not very accurate at determining their relative efficiencies for energy-harvesting. A big reason for this is that the ACS includes power losses in the substrate, rings, and diodes, whereas the experimental efficiency measures only power delivered to the load. Furthermore, for peaks that emerge at higher frequencies, the simulations tend to overestimate the frequency at which they occur. This is probably because the diode model used for the simulation assumes a constant junction capacitance instead of a frequency-dependent one, like found in actual diodes. Since the resonance frequency depends on the capacitance, the observed peaks are shifted.

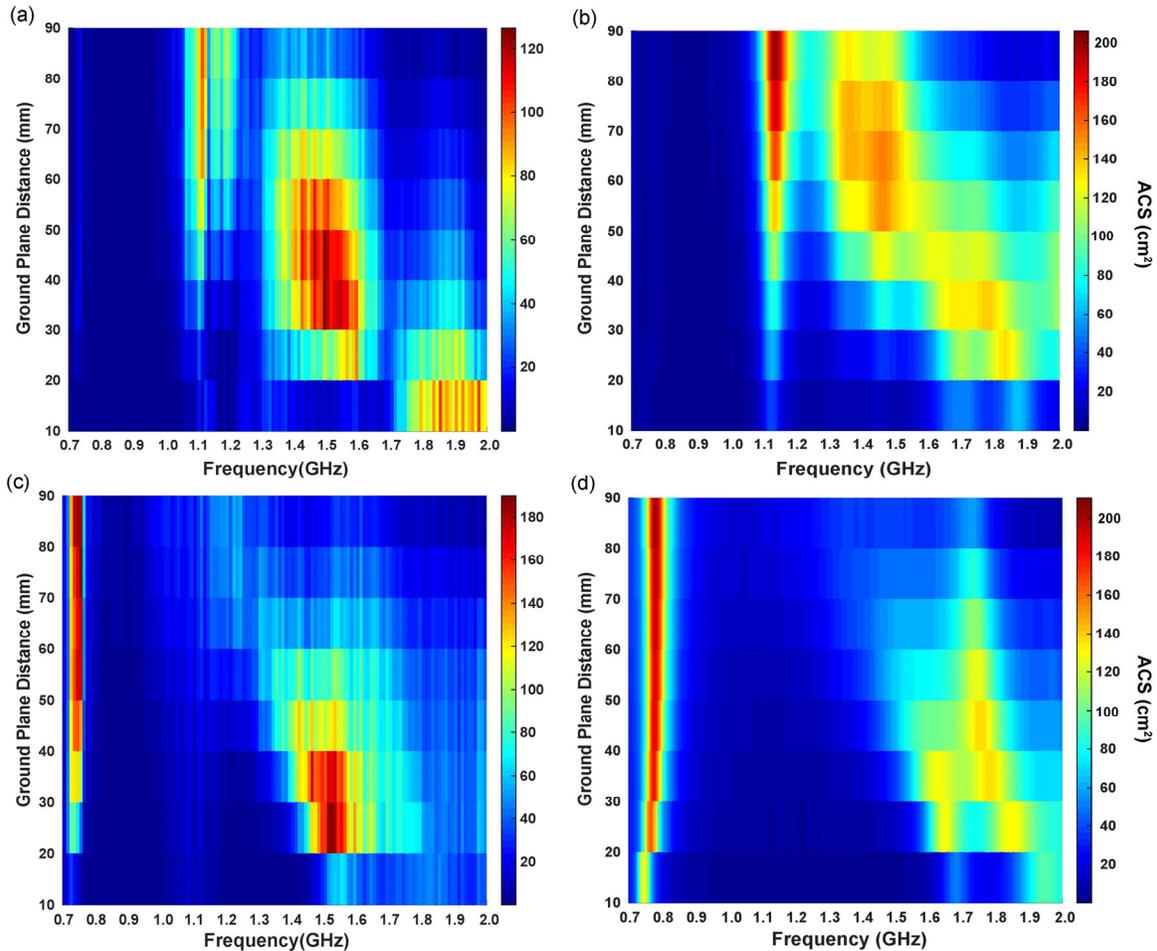

*Figure 3: Ground plane distance dependence. The two polarization angles were chosen to correspond to the minimum and maximum efficiencies for the absorption peak near 0.75 GHz. Efficiency (a) and ACS (b) for a 45 degree polarization. Efficiency (c) and ACS (d) for a 135 degree polarization.*

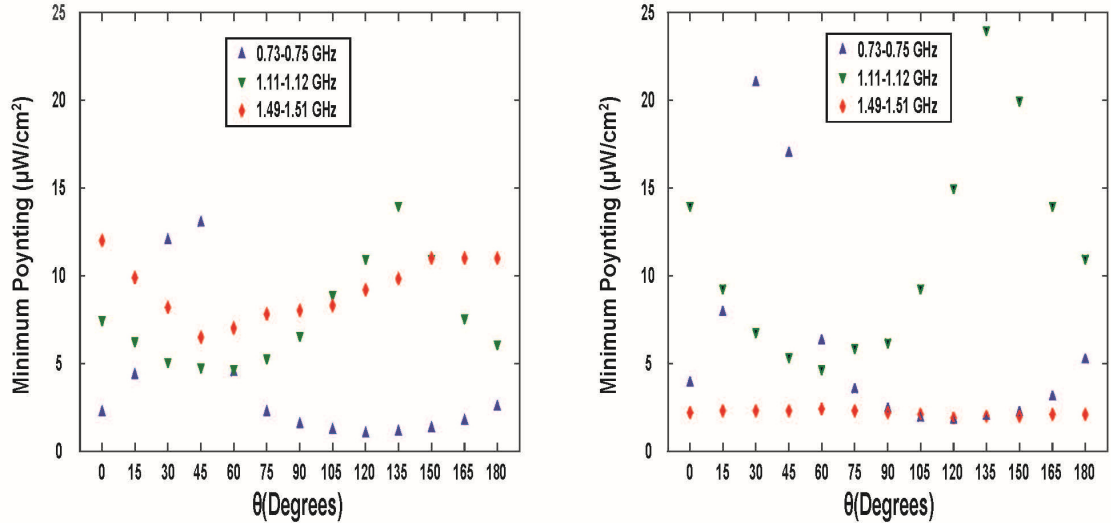

*Figure 4: Minimum power density required to deliver 100 µW to a load as a function of polarization angle. Left: No ground plane. Right: 30 mm ground plane.*

As an additional way of characterizing the performance of the rectenna, particularly for low-power applications, the minimum incident power density required to deliver 100 µW to the load was measured as a function of polarization angle (figure 4). Without the ground plane, the polarization-independence is not very good, but the 0.75 GHz peak is at its most efficient value, being capable of powering a device with only about 1 µW/cm$^2$ of ambient power density available at a 135 degree polarization angle. With the ground plane at 30 mm, the polarization-independent region exhibits excellent uniformity, and is capable of powering a device with only 2.5 µW/cm$^2$ for all polarization angles. The 0.75 GHz peak, while not as efficient with the ground plane present, can still power a device with only 2 µW/cm$^2$ at a 135 degree polarization angle. The weaker peak near 1.1 GHz requires at least 5.0 µW/cm$^2$ and performs slightly better without the ground plane than it does with a cavity length of 3 cm.

**Discussion** The four-ring unit cell was expected to yield polarization-independence due to its rotational symmetry and because the cross-shaped intersection of the four rings interacts with both components of the electric field for arbitrary polarization angles of a normally-incident wave. While the "frame" around the rectenna was intended to simply act as a means for combining the DC for each of the current-paths through the sample, it turned out to be a beneficial component of the design in other ways. First, the frame adds additional effective area for the antenna to capture energy. Second, both the 0.75 GHz and 1.12 GHz resonances would not occur from the 4-ring unit cell alone. Simulations of the surface currents using *CST Microwave Studio* (figure 5) indicate that those resonances are caused by current flowing into and out of the ring structure from the frame. Experimental measurements and simulations using only a single cell enclosed by a frame do not produce a polarization-independent peak, only a dipole-like one. Lastly, when enough unit cells are included to produce the polarization-independent region, the frame is partly responsible for the broad bandwidth because it enables closely-spaced overlapping resonances due to multiple current-paths of similar optical length. Hypothetically, the inclusion of more unit cells should enhance the performance of the polarization-independent modes relative to the dipole-like modes as additional cells should reduce the overall influence of the frame.

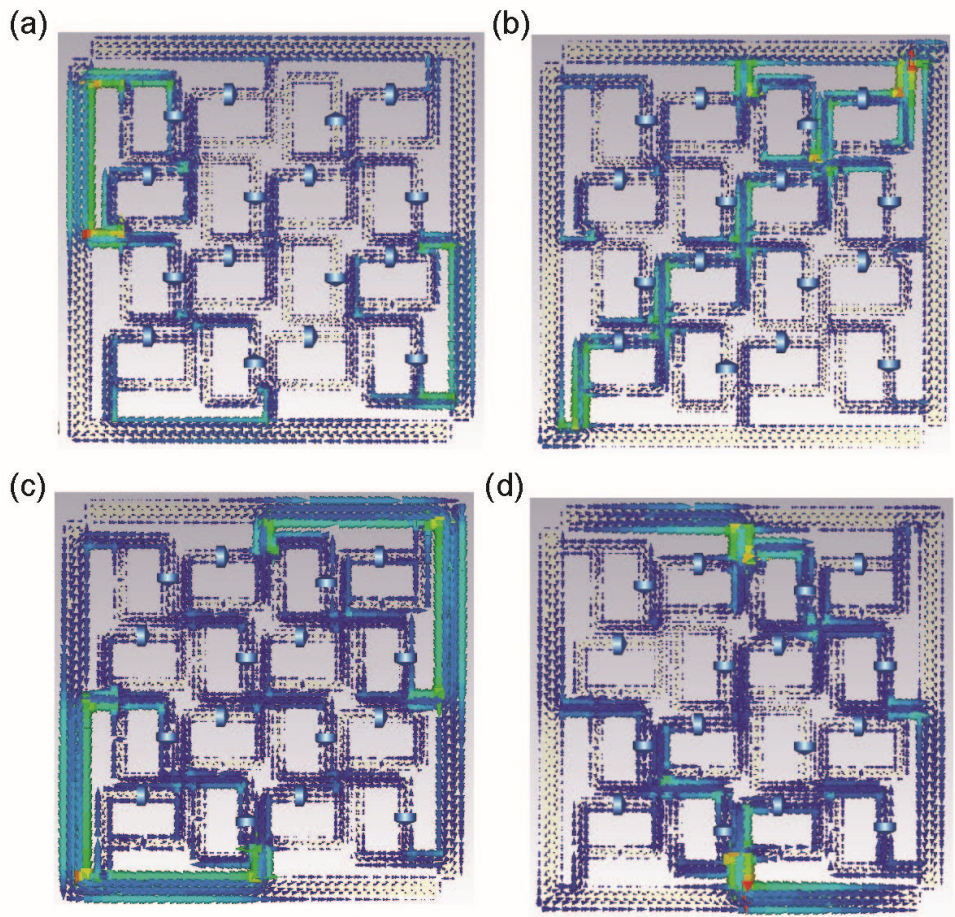

*Figure 5: Simulated current modes for absorption peaks with corresponding experimental frequency reported in parenthesis: (a) 1.13 GHz (1.11) and 45 degree polarization angle (ccw from right horizontal) (b) 1.68 GHz (1.49) and 45 degrees (c) 0.78 GHz (0.75) and 135 degrees (d)1.62 GHz (1.49) and 135 degrees. Note the slight shift in the absorption frequency of the 1.6 GHz (1.49) peak due to the difference in path lengths.*

There is difficulty in comparison of previous results between groups because there are a variety of definitions for efficiency found in the literature, and often times the efficiency being characterized is that of individual components of the energy-harvesting system (antenna efficiency, rectification efficiency, power management circuit efficiency, etc.) instead of the overall efficiency of the system. The antenna efficiency, in particular, is the major source of disagreement as some groups base their definition of efficiency off of the effective area of the antenna instead of the physical area[20,22]. This results in a more traditional definition of efficiency that will not exceed 100%, but we prefer to define efficiency in terms of the geometric area instead for a couple of reasons. The first reason is that theoretically determining the effective area of an antenna containing nonlinear elements is difficult to do and will likely not be very accurate. Measuring the geometric area for planar antennas is usually straightforward. Second, in order to fully characterize the performance of the antenna, a comparison between the effective area and the geometric area would still need to be made. This information is essentially already included when the geometric area is used to define efficiency. Although we are not alone in using the geometric area to define efficiency[1,2,31], we appear to be the first group to report efficiencies over 100%.

There are a large number of designs presented in the literature for MPAs and antennas that exhibit polarization-independence, but very few of these have been adapted with rectification systems which are also supported by experimental energy-harvesting data (some designs have been tested using resistors or else only simulation results using equivalent circuit models are presented). Fewer still have been tested at power densities that are low enough that could be considered ambient (< 50 µW/cm²). Popović et al.[1] and

|  | Popović (1)[1] | Popović (2) | Popović (3) | Kuhn[3] | This Work (Dipole) | This Work (PI) |
|---|---|---|---|---|---|---|
| **Antenna Type** | Yagi-Uda | Patch Array | Spiral Array | Multi-Band Dipole | MPA | MPA |
| **Frequency (GHz)** | 0.915, 2.45 | 1.96 | 2.0-18.0 | 0.9,1.8,2.1,2.4 | 0.75 | 1.49 |
| **Size (cm²)** | 45-80?[a] | 616 | 340 | 100+[b] | 54 | 54 |
| **Efficiency Benchmark**[c] | 40.7%, 56.2% at 1 µW/cm² | 43% at 15 µW/cm² | 20% at 6.2 µW/cm², 2.3% at 0.62 µW/cm² | 30% at 0.1 µW/cm² | 130% at 1 µW/cm², 230% at 10 µW/cm² | 36% at 1 µW/cm², 167% at 10 µW/cm² |
| **$S_{min}$ (µW/cm²)[d]** | < 2 | < 5 | 3 | < 2 | 1 | 2.5 |
| **Polarization** | Linear | Dual-Linear | Circular | Dual-Linear?[e] | Linear | Dual-Linear |
| **Ground Plane** | No | Yes | No | Yes | No | Yes |

Table I

[a] The specific dimensions were not reported and so this is an estimate derived from the efficiency and the amount of power harvested

[b] The area of the antenna is 100 cm², but a larger ground plane of unreported area is also included in the design

[c] These results are not always the highest efficiencies reported for each design, because values for power densities >50 µW/cm² are omitted from this table. Popović uses the same definition for efficiency as this work, so a more direct comparison can be made. It is more difficult to compare with Kuhn, because it is unclear how they determined their antenna size for purposes of measuring efficiency. Additionally, Kuhn's antenna is designed to harvest from up to four frequencies simultaneously and so the performance will be better out in the field than is suggested by its efficiency value

[d] Minimum amount of power density needed to deliver 100 µW/cm² to a load. In most cases, these values were not directly reported for each design. Instead this has been estimated from the efficiency reported and the antenna size

[e] The authors did not make any comments as to the polarization-dependence, but judging from its appearance it is probably *dual-linear*

Kuhn, Lahuec, Seguin, and Person[3] have produced results that appear to be the most comparable to those presented in this work and so are compared with our design in terms of efficiency, polarization-dependence, and ability to deliver power to a load at low power densities (table1). It is important to note that the designs of Popović and Kuhn feature power-management circuits, so in that sense they may be more "field-ready" than the design presented here.

Based on reported efficiencies and comparisons with other designs (some designs exhibiting polarization independence, others not), the design presented in this paper appears to be the most efficient low-power energy-harvesting device to date (for the 0.75 GHz peak) while also being the most efficient polarization-independent design. Additionally, the polarization-independent region is fairly broad and exhibits limited tunability. Furthermore, this design is relatively compact in size (54 cm$^2$) compared to others which harvest similar amounts of power. Power density measurements reported by Visser, Reniers and Theeuwes[8] indicate that in some cases there is enough power available for harvesting within 50 m of a cell-phone tower that this device could potentially continuously power a 100 µW device, although it may need to be rescaled to match frequencies. In practice, a closer distance would likely be required due to fluctuations in output from the tower and also because the design does not yet have a power management circuit to compensate for the non-sinusoidal signals.

There are a few simple ways the design could readily be improved. Work by Kuhn, Seguin, Lahuec, and Person[32] indicates that a MSS20-141 Schottky diodes are superior to SMS-7630 diodes for low power. Additionally, a Rogers TMM10i dielectric substrate[20] could be used. It has a lower loss tangent and higher index of refraction than the FR4 substrate used in this design. This could improve efficiency by reducing losses and the device could be made more compact by filling the cavity with the substrate, which would reduce the ground plane spacing needed for optimal efficiency (although this would introduce some power loss). Additionally, there may be alternate ways to orient the diodes that would improve the efficiency or the uniformity of the polarization-independence. An idea for further exploration would be testing a design with more unit cells contained within the frame to see if the polarization-independent modes would be enhanced and if the bandwidth would increase as more current paths became available. Lastly, although the 4-ring unit-cell was created with energy-harvesting in mind, the electrical connections created by the overlapping cells might result in useful properties that would be desired for other applications.

## **Conclusion**

A multi-resonant RF energy-harvesting rectenna based on a metamaterial perfect absorber with efficiencies exceeding 100% has been presented. It possesses an absorption peak for linear polarizations at 0.75 GHz featuring efficiencies around 230% at 10 µW/cm$^2$ and 130% at 1 µW/cm$^2$, and also features a broader polarization-independent region at 1.4-1.7 GHz (with 167% and 36% maximum efficiencies at those same respective power densities) for which the absorption frequency can be modestly tuned by adjusting the distance between the ground plane and the rectenna. Under ideal conditions, 100 µW of harvested power could be delivered to a device with only 1 µW/cm$^2$ of available power density for the linear mode, and 2.5 µW/cm$^2$ for the polarization-independent mode. These power densities can be found within 50 m of some cell-phone towers.